# A nested loop for simultaneous model topology screening, parameters estimation, and identification of the optimal number of experiments: Application to a Simulated Moving Bed unit

Rodrigo V. A. Santos[a,b,c], Carine M. Rebello[a], Anderson Prudente[a], Vinicius V. Santana[b,c], Ana M. Ribeiro[b,c], Alírio E. Rodrigues[b,c], José M. Loureiro[b,c], Karen V. Pontes[a*], Idelfonso B. R. Nogueira[b,c*]

[a]Programa de Pós-Graduação em Engenharia Industrial (Industrial Engineering Program), Escola Politécnica (Polytechnique Institut), Universidade Federal da Bahia, 40210-630, Salvador (BA), Brazil.

[b] LSRE-LCM - Laboratory of Separation and Reaction Engineering – Laboratory of Catalysis and Materials, Faculty of Engineering, University of Porto, Rua Dr. Roberto Frias, 4200-465 Porto, Portugal

[c] ALiCE - Associate Laboratory in Chemical Engineering, Faculty of Engineering, University of Porto, Rua Dr. Roberto Frias, 4200-465 Porto, Portugal

* To whom correspondence should be addressed: E-mail: karenpontes@ufba.br; idelfonso@fe.up.pt

# Abstract

Simulated Moving Bed (SMB) chromatography is a well-known technique for the resolution of several high-value-added compounds. Parameters identification and model topology definition are arduous when one is dealing with complex systems such as a Simulated Moving Bed unit. Moreover, the large number of experiments necessary might be an expansive-long process. Hence, this work proposes a novel methodology for parameter estimation, screening the most suitable topology of the model's sink-source (defined by the adsorption isotherm equation) and defining the minimum number of experiments necessary to identify the model. Therefore, a nested loop optimization problem is proposed with three levels considering the three main goals of the work: parameters estimation; topology screening by isotherm definition; minimum number of experiments necessary to yield a precise model. The proposed methodology emulated a real scenario by introducing noise in the data and using a Software-in-the-Loop (SIL) approach. Data reconciliation and uncertainty evaluation add robustness to the parameter estimation adding precision and reliability to the model. The methodology is validated considering experimental data from literature apart from the samples applied for parameter estimation, following a cross-validation. The results corroborate that it is possible to carry out trustworthy parameter estimation directly from an SMB unit with minimal system knowledge.

## Nomenclature

| | |
|---|---|
| $b_{A/B}$ | Adsorption parameter |
| $b_{A|B}^{(1)}$ | Adsorption parameter |
| $b_{A|B}^{(2)}$ | Adsorption parameter |
| $c_{A|B}$ | Concentration of enantiomers A and B in the fluid phase |
| $D$ | Axial dispersion coefficient |
| $f_{A|B}$ | Adsorption isotherm equation |
| $F_{obj}$ | Objective function value |
| $f(\mathbf{x})$ | Gaussian probability density function |
| $k$ | Intraparticle mass transfer coefficient |
| $K_{A|B}$ | Adsorption isotherm parameter |
| $L$ | Length of one TMB/SMB column |
| $N_{En}$ | Number of enantiomers |
| $N_{Exp}$ | Number of experiment series |
| $N_{FR}$ | Number of flowrates |
| $N_I$ | Number of iterations |
| $N_{Mea}$ | Number of measurements |
| $N_{OV}$ | Number of operating variables |
| $N_{Pt}$ | Number of particles |
| $N_s$ | Number of columns per section in an SMB unit |
| $N_{SCP}$ | Number of synthetic concentration-points taken in an SMB unit |
| $Pe$ | Peclet parameter, $Pe = (vL)/D$ |
| $q_{A|B}$ | Average concentration of enantiomers A and B in the adsorbed phase |
| $q_{A|B}^*$ | Adsorbed-phase concentration in equilibrium with $c_{A|B}$ |
| $q_S$ | Adsorption isotherm parameter |
| $q_s^{(1)}$ | Adsorption isotherm parameter |
| $q_s^{(2)}$ | Adsorption isotherm parameter |
| $t$ | Time variable |
| $t^*$ | Switching time |
| $u_{c_{A|B}}$ | Uncertainty propagation of $c_{A|B}$ |
| $u_{q_{A|B}}$ | Uncertainty propagation of $q_{A|B}$ |
| $\mathbf{V_y}$ | Square matrix of PSO variance |
| $Var$ | Variance |
| $x$ | Dimensionless variable of length, $x = z/L$ |
| $\mathbf{x}$ | Vector of 30 000 random variables following a Gauss distribution |
| $\mathbf{X}$ | Vectors of the parameters in the uncertainty propagation equations |
| $z$ | Axial coordinate |

### Greek Symbols

| | |
|---|---|
| $\alpha$ | Mass transfer parameter, $\alpha = (kL)/v_s$ |
| $\alpha_{cov}$ | Coverage probability |
| $\gamma\|\gamma^*$ | Ratio between fluid and solid interstitial velocities, $\gamma = v/v_s$ and $\gamma^* = v^*/v_s$ and in TMB and SMB model, respectively |
| $\varepsilon$ | Bed Porosity |
| $\zeta$ | Set of synthetic concentration-points |
| $\theta$ | Dimensionless variable of time, $\theta = t/\tau_s$ |
| $\boldsymbol{\theta}_{PP}$ | New parameter population |
| $\boldsymbol{\theta}_{PSO}$ | Matrix of PSO parameters |
| $\boldsymbol{\lambda}$ | Set of parameters |
| $\lambda_{min\|max}$ | Lower and upper bound of each parameter in $\boldsymbol{\lambda}$ |
| $\mu$ | Gauss mean |
| $\sigma$ | Standard deviation |
| $v\|v^*$ | Interstitial fluid velocity in TMB and SMB model, respectively |
| $v_s$ | Solid interstitial velocity |
| $\boldsymbol{\xi}$ | Set of operating variables |
| $\tau_s$ | Solid space time in a section of a TMB unit, $\tau_s = L/v_S = N_s t^*$ |
| $\boldsymbol{\phi}_{N,np}$ | Matrix of kept PSO population |
| $\chi$ | Set of flowrates |

**Subscript and Superscript**

| | |
|---|---|
| $A\|B$ | Pair of enantiomers |
| $e$ | Experimental |
| $E$ | Eluent stream in SMB/TMB |
| $F$ | Feed stream in SMB/TMB |
| $F_{test}$ | Fisher-Snedecor test |
| $FR$ | Flowrate |
| $i$ | Parameters in the uncertainty propagation equations |
| $j$ | TMB/SMB section |
| $N$ | Number of kept PSO candidates |
| $nc$ | Number of parameters for the calculation of $c_{A\|B}$ |
| $np$ | Number of parameters to be estimated in each adsorption isotherm equation |
| $nq$ | Number of parameters for the calculation of $q_{A\|B}$ |
| opt | Optimum point |
| $p$ | Prediction |
| $R$ | Raffinate stream in SMB/TMB |
| $Rec$ | Flowrate for data reconciliation |
| $X$ | Extract stream in SMB/TMB |

# 1. Introduction

The modeling of a system can be divided into two steps, model definition and parameter estimation. During the model definition, one might face the choice of characterizing an unknown phenomenon represented in the sink-source term of the model. Hence, the nature of the model, its topology, is usually defined before the parameters estimation step, which is a bias introduced in the system. On the other hand, parameter identification of systems that might require a large number of experiments might be an expansive-long process. These problems become evident when one is dealing with complex systems such as Simulated Moving Bed (SMB) chromatography.

SMB is used in different areas of industry such as pharmaceutical, fine chemistry, and biotechnology. Usually, it is applied to promote the separation of complex systems. It has the advantage of being environmentally friendly, cheaper, safer, and faster than other mechanisms. The system comprises two inlet streams: eluent ($E$) and the mixture of compounds to be separated ($F$), and two outlet streams for removal of the more and less retained compounds in the extract ($X$) and raffinate ($R$), respectively. Compounds are separated by the cyclical and periodic movement of the inlet and outlet streams in the system, simulating a counter-current flow of solid and liquid phases as in a True Moving Bed (TMB).[1–5] During the SMB operation, one of the compounds is preferentially adsorbed in the solid phase, whereas the other tends to remain in the liquid phase.

The SMB operation is usually associated with high value-added products, such as enantiomers. Therefore, the process requires tight control of quality variables, which can be well managed if assisted by phenomenological models. The sink-source term of the SMB model is defined by the adsorption leveraged in the system. The mathematical law that describes the adsorption equilibrium is an isotherm equation that expresses the equilibrium concentrations between the liquid and solid phases. According to IUPAC, type 1 adsorption isotherm equations have a good representation of SMB liquid separation in microporous adsorbents [6], which are the focus of this work. When modeling the SMB using type 1 isotherm, different adsorption isotherm equations have been proposed in the literature: Linear (L), Langmuir (LG), Linear + Langmuir (LLG), and Bi-Langmuir (BLG) isotherms.

Several compounds can be separated by SMB, and for each system, a different adsorption isotherm equation might be more suitable. Pais (1999)[7] evaluated the aforementioned adsorption isotherm equations and adopted the Bi-Langmuir isotherm as the most suitable equation for separating 1,1'-bi-2-naphthol enantiomers in 3,5-dinitrobenzoyl phenylglycine bonded to silica gel. Mun (2012, 2016)[8,9] studied the separation region in SMB chromatography applying linear isotherms for binary separation of mixtures of L-phenylalanine and L-tyrosine, and acetic acid, respectively. Graça, Pais and Rodrigues (2015)[10] also adopted a linear adsorption isotherm for ternary mixtures such as oligosaccharide in the pseudo-simulated moving bed process, known as JO process. Aniceto, Cardoso and Silva (2016)[11] apply a modified Linear + Langmuir isotherm for the separation of (S,S)-(−) and (R,R)-(+)-trans-stilbene oxide enantiomers in Chiralcel OD adsorbent. Arafah et al. (2016, 2018, 2020)[12–14] studied the separation of nadolol stereoisomers using Chiralpak IA and adopted a Linear + Langmuir isotherm to describe the adsorption behavior. Huthmann and Juza (2005)[15] also apply the Linear + Langmuir isotherm for less common applications of simulated moving bed chromatography in the pharmaceutical industry. Cunha (2021)[16] studied the separation via SMB of a medicine against schistosomiasis, praziquantel, and considered a Langmuir isotherm. Shih et al. (2019)[17] propose using Langmuir isotherm for the case of $CO_2$ adsorption in a metal-organic framework for environmental purposes. These works do not make an insightful analysis of which isotherm is best suited for a given system. The choice is usually made based on parameters of a certain isotherm already available in the literature. Another possibility is to measure the average concentrations in equilibrium between adsorbed and fluid phases by breakthrough experiments and then they are adjusted to the isotherm equation that better suits a particular system. Different isotherm curves are then adjusted in a trial-and-error procedure until a good quality of fit is achieved.[7] However, a rigorous identification of the adsorption equation is seldom done.

Regardless of the most suitable adsorption isotherm, the SMB model requires parameters associated with diffusion, mass transfer, and adsorption phenomena. Traditionally SMB parameter estimation is carried out by the "Step by Step" approach that might involve tracer tests, breakthrough and pulse experiments, followed by validation in an SMB unit.[18] This last step is usually carried out without much statistical rigor, except for a few works in the literature.[5,19–25] If the parameters do not provide a good model representation, the model must be rearranged by redefining its adsorption isotherm equation in order to set another one that better suits the system. This traditional "Step by Step" model identification has several steps and procedures that make it costly and slow.[18] A previous study proposed a methodology that enables a robust global parameter estimation for the SMB model, based on considerations and previous knowledge from the literature, and, concomitantly, adds reliability to the model through an uncertainty evaluation.[18] Up to the authors' knowledge, the topic of isotherm selection for a generic system, together with a global parameter estimation and uncertainty analysis, has not been thoroughly evaluated in the literature so far. Therefore, this paper proposes a methodology for SMB adsorption isotherm and model identification based on minimal knowledge that it is able to simplify the entire process.

Isotherm selection and parameter estimation depend on experimental data that, inevitably, bears measurement errors that might compromise the quality and reliability of the model prediction. To reduce measurement errors and adjust the collected data, reconciliation of experimental data has to be carried out. Although this is an important issue, it is often overlooked in the literature. Suzuki et al. (2021)[25] addressed this problem in the context of SMB parameter estimation. The authors propose a refinement of SMB model predictions by the sequential approach of Tikhonov regularization, a technique that reduces the effect of measurement errors by applying penalty deviations of parameters based on a reliable reference.[25] The investigation of the quality of data obtained from an SMB plant is fundamental to obtain a more reliable model, therefore data reconciliation is part of the contributions of this work.

Moreover, uncertainty analysis should also be investigated when identifying models, especially in chromatographic models. In the context of the SMB, this problem has received some attention. Yamamoto, Yajima & Kawajiri (2021)[26] proposed a low computational effort Monte Carlo-based algorithm for parameter uncertainty estimation for chromatographic processes, such as SMB, taking glucose and fructose resolution as a case study. Grosfils et al. (2007, 2009, and 2010)[1,19,27], in turn, developed pioneer works of uncertainty analysis for batch experiments in the SMB field, taking its dead volumes into consideration. However, as far as we know, only Santos et al. (2021)[18] approached uncertainty evaluation for a global parameter estimation, and the work proposed herein deepens the subject adding data reconciliation to SMB parameter estimation in order to diminish measurement errors present in a real unit.

This article proposes a new global methodology for model topology identification, parameters estimation and evaluation of the minimum number of experiments necessary to perform the model identification. As a case study, an SMB model is identified, which comprehends adsorption isotherm screening from operational data based on minimal system knowledge. This methodology also intends to consider a real case scenario where data are measured with noise and several sources of uncertainties. Therefore, noise is added to the synthetic data generated. Hence it takes into consideration the data reconciliation and uncertainty evaluation. Thus, considering the uncertainties related to an experimental procedure. A nested-loop-optimization problem is proposed here to achieve the overall goal, which considers the following main point:

- An outer loop for the estimation of the parameters in an overall approach;
- An intermediary loop for model topology identification, isotherm screening;
- An inner economic loop for evaluating the minimum number of experiments necessary to yield a reliable model.
- A data reconciliation term to take into consideration the experimental uncertainties;

The above-listed points can also be highlighted as the main contributions of this work. The methodology validation is done with experimental data from the literature, different from the samples applied for parameter identification to follow a cross-validation approach.

## 2. TMB and SMB Models

TMB and SMB models are described as Partial Differential Algebraic Equation (PDAE), with a total of seven operating variables ($N_{OV}$) for TMB and SMB models: $[c_A^F, c_B^F, t^*, v_{IV}, v_X, v_F, v_E]$ and $[c_A^F, c_B^F, t^*, v_{IV}^*, v_X^*, v_F^*, v_E^*]$, respectively, where $c$ is the concentration, $t^*$ is the switching time, $v$ is the interstitial fluid velocity. Subscript $A$ and $B$ are the two enantiomers, $X$, $F$, $E$ are the Extract, Feed and Eluent streams, and $IV$ is the recycling stream. Superscript $F$ designates the Feed stream, * in $v$ designates the SMB model.

In this work, the Bi-Langmuir-isotherm SMB model is used as a virtual plant for the purpose of emulating a real SMB unit. In contrast, the TMB model is used to estimate model parameters, due to its easier implementation. Component and global mass balances for TMB and SMB models are presented in Table 1 and Table *2*, respectively. There is no solid phase movement in the SMB model and the two models are equivalent by keeping constant the fluid velocity relative to the solid interstitial velocity, as given by:

$$v_j^* = v_j + v_s \tag{1}$$

where $v_s$ is the solid interstitial velocity where $\theta$ is the dimensionless variable of time, $\gamma$ is the ration between fluid and solid interstitial velocities, $x$ is the dimensionless variable of length, $\varepsilon$ is the bed porosity, $q$ is the average concentration of enantiomers, $q^*$ is the adsorbed phase concentration in equilibrium with $c$.

Table 1: Component mass balance for TMB|SMB models.

| | |
|---|---|
| $\frac{\partial c_{A\|B}}{\partial \theta} = \gamma\|\gamma^* \left\{ \frac{1}{Pe} \frac{\partial^2 c_{A\|B}}{\partial x^2} - \frac{\partial c_{A\|B}}{\partial x} \right\} - \frac{(1-\varepsilon)}{\varepsilon} \alpha (q^*_{A\|B} - q_{A\|B})$ | (2) |
| $\frac{\partial q_{A\|B}}{\partial \theta} = \frac{\partial q_{A\|B}}{\partial x} + \alpha(q^*_{A\|B} - q_{A\|B}) \| \frac{\partial q_{A\|B}}{\partial \theta} = \alpha(q^*_{A\|B} - q_{A\|B})$ | (3) |
| $q^*_{A\|B} = f_{A\|B}(c^F_A, c^F_B)$ | (4) |

Table 2: Global mass balance for TMB|SMB models.

| | | |
|---|---|---|
| Eluent (E): | $v_I = v_{IV} + v_E \| v^*_I = v^*_{IV} + v^*_E$ | (5a) |
| Extract (X): | $v_{II} = v_I - v_X \| v^*_{II} = v^*_I - v^*_X$ | (5b) |
| Feed (F): | $v_{III} = v_{II} + v_F \| v^*_{III} = v^*_{II} + v^*_X$ | (5c) |
| Raffinate (R): | $v_{IV} = v_{III} - v_R \| v^*_{IV} = v^*_{III} + v^*_R$ | (5d) |

Both TMB and SMB models have kinetic and mass transfer parameters ($Pe$ and $\alpha$) and adsorption isotherm parameters ($K_A$, $K_B$, $q_S$, $b_A$, $b_B$, $q_S^{(1)}$, $q_S^{(2)}$, $b_A^{(1)}$, $b_A^{(2)}$, $b_B^{(1)}$, and $b_B^{(2)}$) depending upon the adopted adsorption isotherm equation, $f_{A|B}(c^F_A, c^F_B)$. Linear (L), Langmuir (LG), Linear + Langmuir (LLG), and Bi-Langmuir (BLG) equations (shown in Table 3) present good representation for type 1 adsorption isotherms and, therefore, are candidates to be the function $f_{A|B}$ in Eq. (4).

Table 3: Examples of adsorption isotherm equations for type 1 adsorption.

| Adsorption isotherm | Equations $f_{A\|B}(c^F_A, c^F_B)$ | |
|---|---|---|
| L | $q^*_A = K_A c^F_A$ | (6a) |
| | $q^*_B = K_B c^F_B$ | (6b) |
| LG | $q^*_A = \dfrac{q_S b_A c^F_A}{1 + b_A c^F_A + b_B c^F_B}$ | (7a) |
| | $q_B^* = \dfrac{q_S b_B c_B^F}{1 + b_A c_A^F + b_B c_B^F}$ | (7b) |
| LLG | $q_A^* = K_A c_A^F + \dfrac{q_S b_A c_A^F}{1 + b_A c_A^F + b_B c_B^F}$ | (8a) |
| | $q_B^* = K_B c_B^F + \dfrac{q_S b_B c_B^F}{1 + b_A c_A^F + b_B c_B^F}$ | (8b) |
| BLG | $q_A^* = \dfrac{q_S^{(1)} b_A^{(1)} c_A^F}{1 + b_A^{(1)} c_A^F + b_B^{(1)} c_B^F} + \dfrac{q_S^{(2)} b_A^{(2)} c_A^F}{1 + b_A^{(2)} c_A^F + b_B^{(2)} c_B^F}$ | (9a) |
| | $q_B^* = \dfrac{q_S^{(1)} b_B^{(1)} c_B^F}{1 + b_A^{(1)} c_A^F + b_B^{(1)} c_B^F} + \dfrac{q_S^{(2)} b_B^{(2)} c_B^F}{1 + b_A^{(2)} c_A^F + b_B^{(2)} c_B^F}$ | (9b) |

Each isotherm in Table 3 has its own feature and is more suitable for a certain system. Linear isotherm has the simplest formula with only two parameters to be estimated; it is limited for low concentration systems where a linear phase equilibrium concentration is observed.[28] Langmuir isotherm, which considers the component competition for chiral sites, presents a good representation with only three parameters. However, it demands a solid phase with a constant selectivity factor, which is not common in SMB applications. Linear + Langmuir adds two linear parameters to the Langmuir formula.[14,15,28] Linear + Langmuir isotherm presents nonselective and selective terms, $K$, and Langmuir parameters respectively.[29] Finally, Bi-Langmuir isotherm has the most complex and complete formula with six parameters; it describes adsorption in the presence of two distinct sites, which is common for enanatioresolution.[28,30,31]

## 3. Design of experiments

Regardless of the adopted adsorption isotherm equations, the parameters should be estimated with statistical-based experiments within wide limits in order to prevent the model predictions from going to a non-representative zone. In other words, experiments should comprehend different scenarios, especially in the absence of previous knowledge, and the minimum number of experiments to be carried out should be carefully set. In the proposed methodology, Latin Hypercube Sampling (LHS) generates four matrices $\boldsymbol{\xi}_i$, with dimension $N_{Exp} \times N_{OV}$, where $N_{Exp}$ is the number of experiments in each matrix that ranges from 5 to 20 in increments of 5 and $N_{OV}$ is the number of operating variables, $[t^*, v_{IV}^*, v_X^*, v_F^*, v_E^*]$. These four series are generated in order to assess how the number of experiments affects the screening of the adsorption equation. It is worth noticing that $c_A^F$ and $c_B^F$ are also operating variables, however they are kept fixed in 5 g.L$^{-1}$ and, therefore, they are not included in the LHS.

The proposed LHS samples are five-dimensional matrices with well-dispersed and non-overlapping points. The sampling points are shown in two-dimensional sectors of the hyperdimensional landscape in order to show how samples are disposed to one another. Operating variables of the samples $\boldsymbol{\xi_i}$ and their dispersions are presented in Figure 1. The operating variables vary within lower and upper bounds (LB and UB), as shown in Table 4. $\boldsymbol{\xi_1}$ to $\boldsymbol{\xi_4}$ samplings are implemented in the virtual plant yielding synthetic data of experimental concentration profiles for compounds A and B. The virtual plant considers a non-linear Bi-Langmuir adsorption parameters estimated by Pais (1999)[7], as shown in Table 5.

Table 4: Upper and lower bonds of operating variables.

|  | LB | UB |
|---|---|---|
| $t^*$ / (min) | 1,0 | 5,0 |
| $v_{IV}^*$ / (mL/min) | 2.5 | 45.0 |
| $v_X^*$ / (mL/min) | 2.5 | 35.0 |
| $v_F^*$ / (mL/min) | 0.5 | 9.0 |
| $v_E^*$ / (mL/min) | 2.5 | 45.0 |

Table 5: Bi-Langmuir adsorption parameters used in the virtual plant.

| | | | |
|---|---|---|---|
| $Pe$ | 2 000 | $b_A^{(1)}$ / (L/g) | 0.0336 |
| $\alpha$ | 36.00 | $b_A^{(2)}$ / (L/g) | 1.00 |
| $q_s^{(1)}$ / (g/L) | 80.04 | $b_B^{(1)}$ / (L/g) | 0.0446 |
| $q_s^{(2)}$ / (g/L) | 0.10 | $b_B^{(2)}$ / (L/g) | 3.00 |

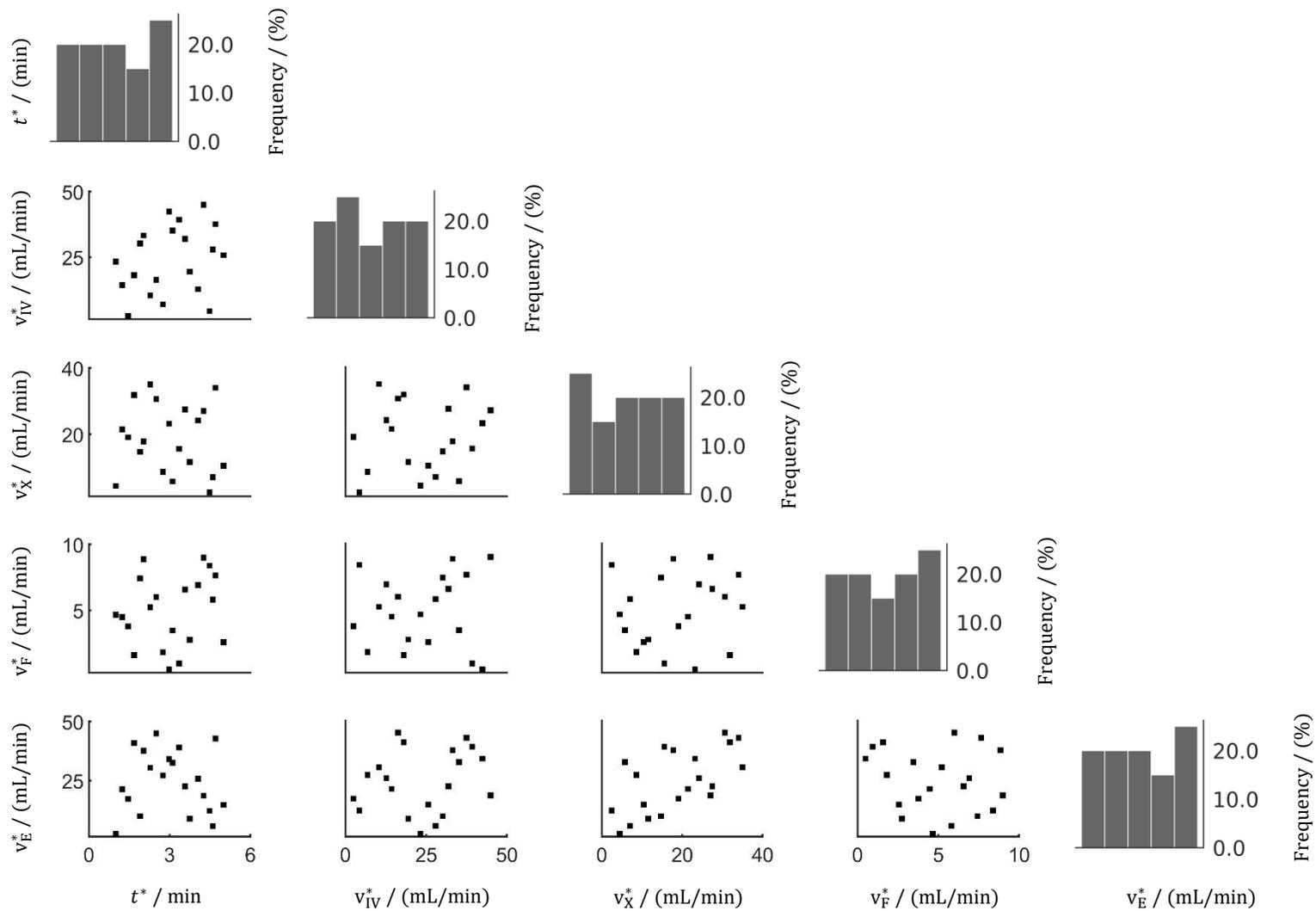

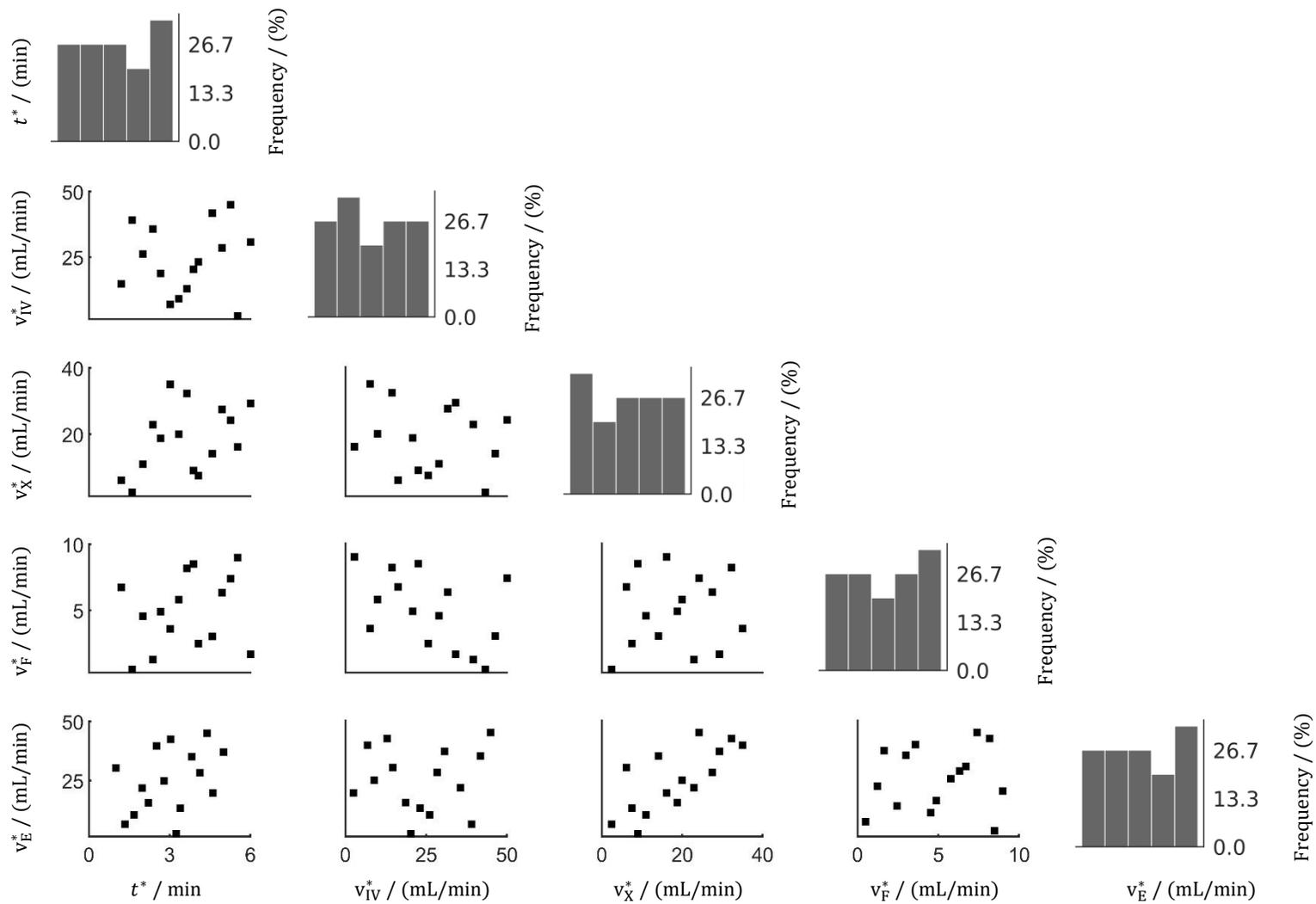

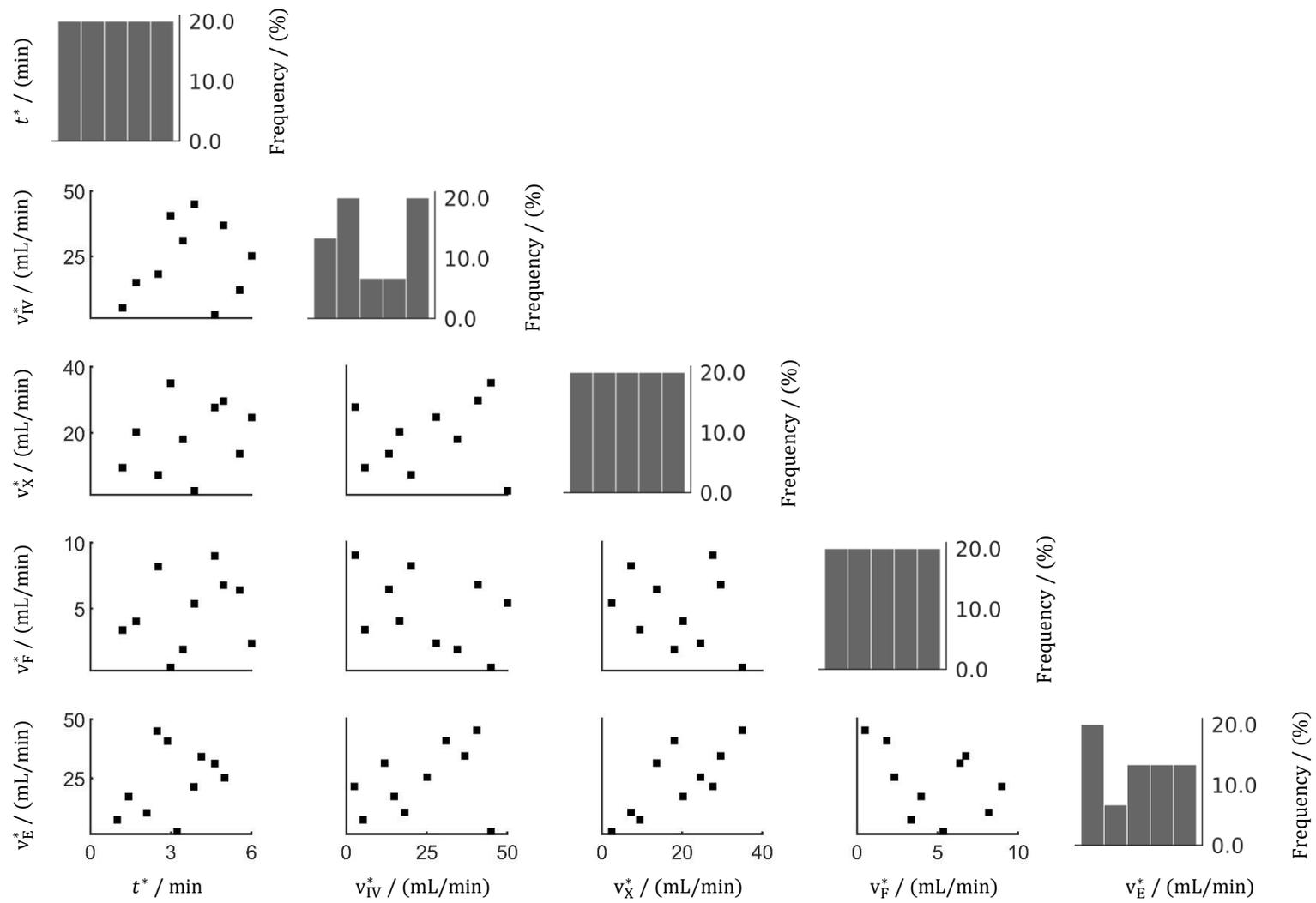

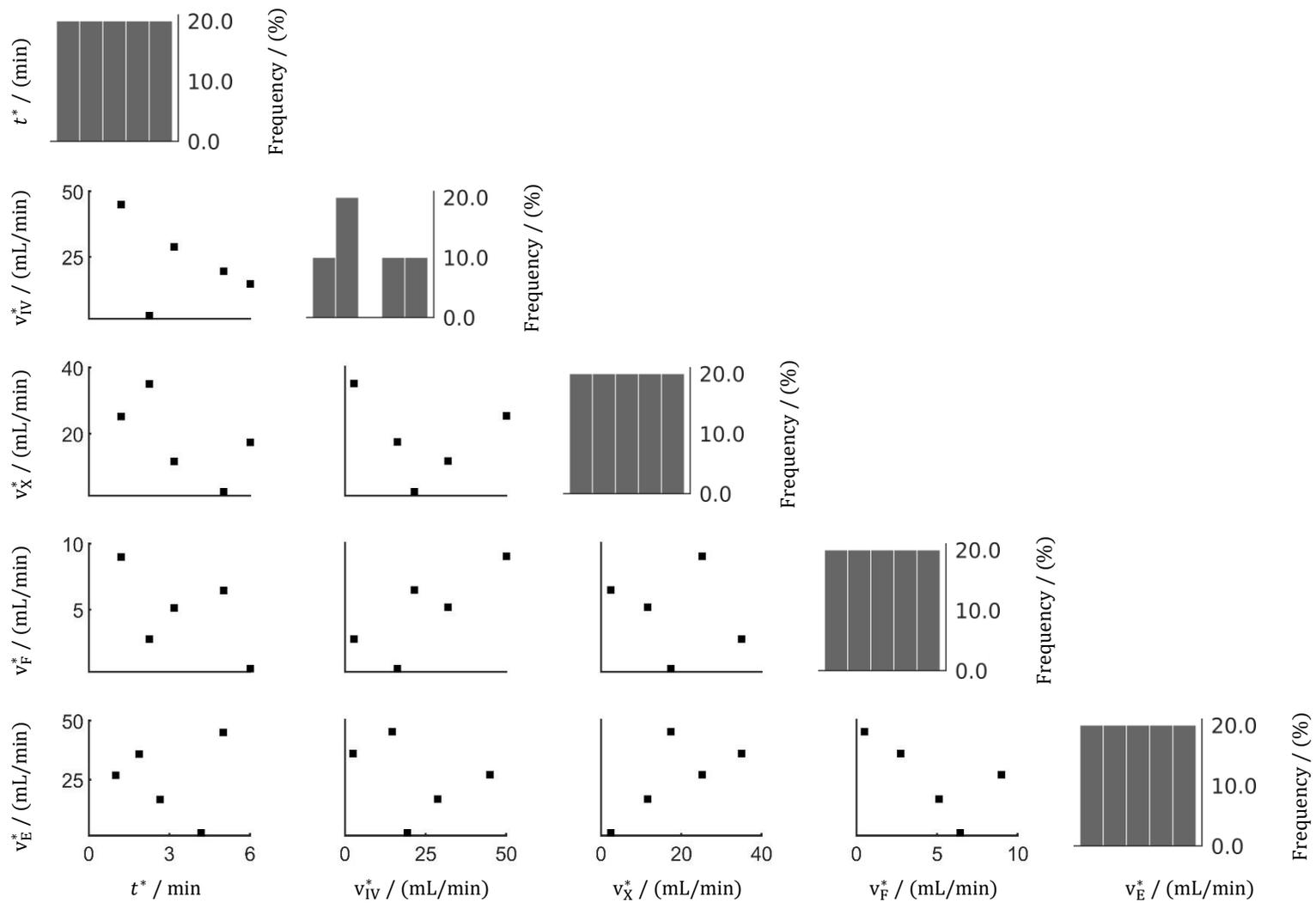

Figure 1: Distributions of operating variables for the four experimental series and their dispersions.

Measurements are taken at different times corresponding to a sampling frequency of 50% of the switching time. Nine synthetic concentration-points, $N_{SCP}$, are taken until the end of the cycle for the two enantiomers (A and B), $N_{En}$, generating a matrix $\zeta_{N_{En} \times N_{SCP}}$. This work adopts the operating variables and model parameters of the SMB virtual plant used in Pais (1999)[7] implemented in the virtual plant whose configuration is 2-2-2-2 (two columns per section). Synthetic data are generated through simulation of the SMB model.

## 4. Data reconciliation and parameter estimation

In chemical processes, control variables as flowrates, concentration, and temperature, are often measured and recorded.[32,33] These data contain measurement errors that compromise the mass and energy conservation.[30] Measurement errors are caused by fluctuations in power supply, network transmission, unexpected changes in the environment, and so on.[32,33] It is impossible to fully eliminate these effects because they are intrinsic to the measurement. These errors may be classified into gross and random errors. The former are related to malfunctioning, lack of calibration, corrosion, etc.; they cause significant variations between the measured and real value and should be avoided with good engineering practices.[34] The latter is related to uncontrollable factors that can be minimized but never nulled. Good measurements present then low disturbance caused by random errors and are absent of gross errors.[34] Random errors are described by gauss curves with low standard deviation.[35] Hence, in this work, gross errors are neglected, and random errors are added to the synthetic data.

The presented methodology emulates experiments in a real SMB unit where measurement noises are present. Therefore, this work simulates random noise in measurements taken under the same condition. Measurement errors are added to each flowrate in $\xi$ ($\chi = [v_{IV}^*, v_X^*, v_F^*, v_E^*]$) in the simulation inlet and to synthetic concentration points in $\zeta$ withing a Gaussian distribution. For each variable and concentration, five measurements, $N_{Mea}$, are taken repetitively in a short time length to simulate the noise of a real experiment. The repeatability procedure together with the noise measurement, provide information for the data reconciliation during parameter estimation and isotherm screening. The added simulated noises in the measurements for the flowrates are within gaussian curves with 0-mean and standard deviation of 26.3%, based on values already applied by Pais (1999)[7] and 1% of experimental concentration average.

Data reconciliation is a technique that adjusts process measurements in a way that collected data satisfy the mass and energy balance constraints of the model.[32,36] It is a constrained least square problem posed to minimize measurement errors whose values are used to define the model parameters that better describe experimental results.[37] The raw plant measurements have imbued errors that are not the solution of a nonlinear system. Data reconciliation is required in model fitting and validation when real data are used in order to correct the model. The international standard ISA-95 states that data reconciliation is important for enterprise-control system integration and accurate data should be provided. Otherwise one might obtain a poor interpretation of the real system.[38] Data reconciliation has been implemented in the methodology proposed here, considering only steady-state TMB data. The technique enables a reduction of the random error effects, this way enhancing the accuracy of the measurements and, concomitantly, the quality of the parameter estimation.

The number of experimental series ($\xi_1$ - $\xi_4$) is used in the inner loop of the nested problem whose goal is to achieve a set of parameters $\lambda$ that minimizes the difference between experimental and predicted enantiomer concentrations along the SMB columns. TMB parameters will be estimated globally by Software-In-the-Loop (SIL) scheme solved by Particle Swarm Optimization (PSO) in the software Matlab® connected to the SMB phenomenological models in gProms®, each isotherm at a time. This optimization method has the advantage of generating historical data that is useful to develop a proper uncertainty evaluation regarding measurements, data, and parameters. Uncertainty evaluation enables comprehension of how reliable the model prediction is through confidence intervals.

The optimization problem proposed here has an objective function designed to incorporate parameter estimation, isotherm screening, and the identification of the minimum number of experiments necessary to achieve a precise model. These are the three loops that the proposed nested loop problem contains. A data reconciliation term is added to the mathematical expression of the objective function.

Hence, the objective function proposed in this work is a nested loop, in which the inner loop evaluates the minimum number of necessary experiments, so considering the economic goals. The intermediary loop is the parameter estimation. The outer loop is related to minimizing the inner problems considering the four possible isotherm equations (Linear, Langmuir, Linear + Langmuir, and Bi-Langmuir) as a discrete decision variable. Figure 2 illustrates the proposed algorithm.

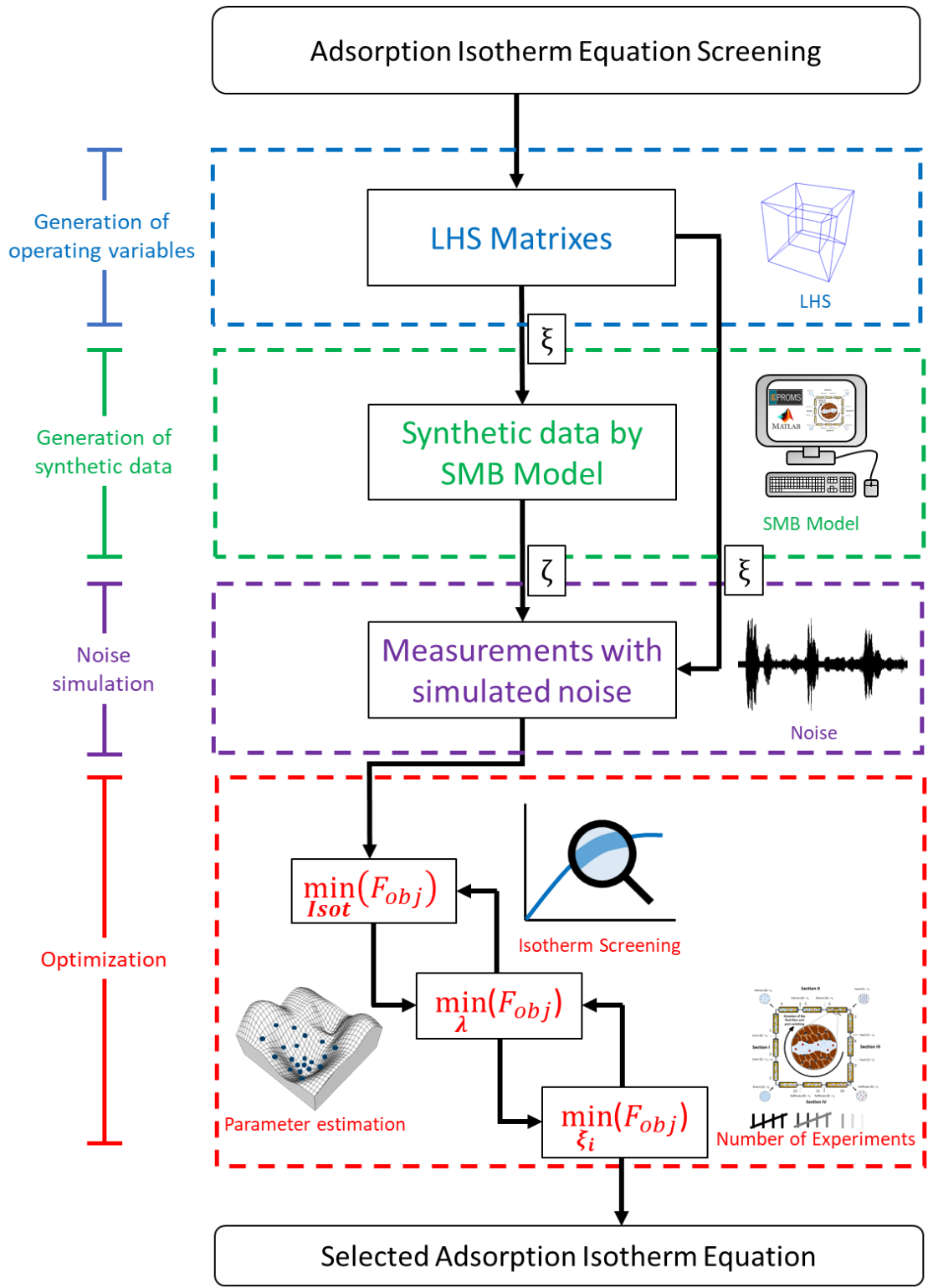

Figure 2: Algorithm for parameter estimation and adsorption isotherm equation screening.

Thus, the optimization problem can be read as a search to find the number of experiments ($\xi_i$), set of model parameters ($\lambda$) and the most suitable isotherm that will maximize the model probability to represent the experimental data fulfilling the imposed constraints. Both data reconciliation and parameter estimation are least square problems. Therefore, they are given by:

$$\min_{Isot}(\min_{\lambda}(\min_{\xi_i} F_{obj}))$$

$$= \sum_{k=1}^{N_{Exp}} \sum_{j=1}^{N_{SCP}} \sum_{i=1}^{N_{Mea}} \left[ \frac{\left(c_{A,i,j,k}^p - c_{A,i,j,k}^{e,k}\right)^2}{Var_A} + \frac{\left(c_{B,i,j,k}^p - c_{B,i,j,k}^e\right)^2}{Var_B} \right]$$

$$+ \sum_{k=1}^{N_{Exp}} \sum_{n=1}^{N_{FR}} \sum_{i=1}^{N_{Mea}} \left[ \frac{\left(\chi_{n,k}^{Rec} - \chi_{i,n,k}^e\right)^2}{Var_{FR}} \right] \quad (10)$$

s.t.:

$$\xi_{min} \leq \xi_i \leq \xi_{max}$$

$$\lambda_{min} \leq \lambda \leq \lambda_{max}$$

$$c_{A,i}^p = \mathbf{f}(\dot{\mathbf{x}}, \mathbf{x}, \lambda, \xi, \mathbf{p}, t)$$

$$c_{B,i}^p = \mathbf{f}(\dot{\mathbf{x}}, \mathbf{x}, \lambda, \xi, \mathbf{p}, t)$$

where $N_{FR}$ is the number of flowrate (4), $Var$ is the variance, $\xi_i$ is an integer variable that varies from 5 to 5, subscript $FR$ is the flowrate and $Rec$ is the flowrate for data reconciliation, superscript $p$ and $e$ designate the prediction and experimental. The first summation is the parameter estimation, subscript $i$ ranges from 1 to $N_{Mea}$, the total number of measurements repetitively taken in each 50% of the time switch (5), $j$ ranges from 1 to $N_{SCP}$, total number of synthetic concentration-points taken (9) in a given experiment. The second summation is the data reconciliation term, in which the subscript $n$ ranges from 1 to $N_{FR}$, the total number of flowrates in the operating variables, $[v_{IV}^*, v_X^*, v_F^*, v_E^*]$, (4).

PSO is used to solve Eq. (10) due to its easy implementation, less computational effort when compared to other methods, and non-dependence of initial guesses when compared to other heuristic methods.[18] Furthermore, PSO is frequently applied for parameter estimation, especially in adsorption cyclic processes, such as SMB. Santos et al. (2021)[18] applied parameter estimation successfully, using PSO for a global SMB parameter estimation, taking advantage of the population generated by the PSO to carry out the uncertainty analysis.

## 5. Uncertainty Evaluation

Well-estimated parameters enable the development of representative models. Nevertheless, the model prediction can yet bear inaccuracies that, without their evaluation, might not portray the actual behavior of a process. Uncertainty evaluation is crucial in order to assess parameter accuracy and to obtain more trustful and precise models. The PSO population provides valuable information about the system uncertainty through the Fisher-Snedecor test without additional calculations.

Parameters that meet Fisher-Snedecor test are used to build the likelihood confidence regions and are kept in matrices $\boldsymbol{\phi}_{N,np}$ that are further used to build the confidence regions,[39] where $N$ is the number of of kept PSO candidates and $np$ is the number of parameters to be estimated in each adsorption isotherm equation. Likelihood confidence regions enable the quantification of parameter uncertainty.[18] This tool ranks all the population and selects the ones with a higher degree of confidence.[40] This technique was used by other authors such as Schwaab and Pinto (2007)[40], Benyahia et al. (2013)[39], Nogueira et al. (2018, 2019a)[41,42] and Santos et al. (2021)[18]. Uncertainties are then identified and propagated to the model prediction with the optimal adsorption isotherm. Usually confidence regions are built considering a normal distribution of the estimated parameters, consequently leading to elliptical regions.[43,44] On the other hand, when using PSO, the matrices $\boldsymbol{\phi}_{N,np}$ indicates the shape of the confidence region. Correlation of parameters is inferred by the ellipse inclination. A round shape ellipse indicates that the parameters present high covariance. The confidence regions enable the propagation of the uncertainty to SMB concentration profiles as well as the selected adsorption isotherm model. The methodology for uncertainty evaluation was originally described by Santos et al. (2021)[18].

The PSO population does not fill the confidence region thoroughly, leaving empty spaces. The use of a larger PSO population in an attempt to fill these empty spaces brings higher computational effort. Therefore, a trade-off between computational effort and regions fulfillment needs to be considered. The vacancies are filled with the region shape information by employing a method based on the Multivariate Law of Propagation of Probability Density Functions (MLPP), using Monte Carlo (MC) simulations, which requires less computational effort than re-evaluating the objective function.[18]

Covariance matrices are used to fill this empty space. A normal gaussian with zero mean ($\mu = 0$) and standard deviation equal to one ($\sigma = 1$) was considered to fill the confidence region. It creates points respecting the ellipse limits ($\alpha_{cov} = 99.9\%$, equivalent to $-3.291 \leq \mathbf{x} \leq 3.291$). New points are generated within the generated ellipse in order to fill the empty space, according to:

$$f(\mathbf{x}) = \frac{1}{s\sqrt{2\pi}} e^{-\frac{1}{2}\left(\frac{\mathbf{x}-\mu}{s}\right)^2} \tag{11}$$

$$\boldsymbol{\theta}_{PP} = \boldsymbol{\theta}_{PSO} + \mathbf{V}_y f(\mathbf{x}) \tag{12}$$

where $f(\mathbf{x})$ is the Gaussian probability density function, $s$ is the standard deviation, $\mu$ is the Gauss mean, $\mathbf{x}$ is a vector of 30 000 random variables, $\boldsymbol{\theta}_{PSO}$ is the matrix of PSO parameters, $\boldsymbol{\theta}_{PP}$ is a new parameter population and $\mathbf{V}_y$ is a square matrix of PSO variance.

Regardless of the screened isotherm equation, the model parameters are assumed to contribute equally to the uncertainty propagation. Parameter uncertainties are propagated according to the Guide to the Expression of Uncertainty in Measurement (GUM) and, based on frequency distribution, parameter uncertainties are computed as type A.[45] Variable uncertainties for the concentrations in the fluid phase and adsorbed phase, $u_c$ and $u_q$, are given by:

$$u_{c_{A|B}} = c_{A|B} \sqrt{\frac{\sum_{i=1}^{nc} u_i^2(\mathbf{X_i})}{\sum_{i=1}^{nc} \mathbf{X_i^2}}} \tag{13}$$

$$u_{q_{A|B}} = q_{A|B} \sqrt{\frac{\sum_{i=1}^{nq} u_i^2(\mathbf{X_i})}{\sum_{i=1}^{nq} \mathbf{X_i^2}}} \tag{14}$$

where $u_{c_{A|B}}$ and $u_{q_{A|B}}$ are the uncertainty propagations of $c_{A|B}$ and $q_{A|B}$, $nc$ and $nq$ are the number of parameters for the calculation of $c_{A|B}$ and the number of parameters for the calculation of $q_{A|B}$ and $\mathbf{X}$ are vectors of the parameters in the uncertainty propagation equations.

## 6. Results

### 6.1. Parameter estimation, adsorption isotherm equation screening, minimum number of experiments and analysis

The nested-loop optimization problem from Figure 2 was solved by a Particle Swarm Optimizer. In this way, it was identified the minimum number of experiments necessary, 15, the most suitable adsorption isotherm, Bi-Langmuir, and the corresponding model parameters, as shown in Table 6.

Table 6: Estimated parameters, screened adsorption isotherms equation and minimum number of experiments for TMB model.

|  | Optimum Point | Bi-Langmuir ($\xi_2$) |
|---|---|---|
|  | Number of experiments | 15 |
| Parameters | $Pe$ | 2 036 |
| Parameters | $\alpha$ | 2.70 |
| Parameters | $q_s^{(1)}$ / (g/L) | 41.34 |
| Parameters | $q_s^{(2)}$ / (g/L) | 74.82 |
| Parameters | $b_A^{(1)}$ / (L/g) | 5.76 |

|  |  |
|---|---|
| $b_A^{(2)}$ / (L/g) | 78.66 |
| $b_B^{(1)}$ / (L/g) | 25.91 |
| $b_B^{(2)}$ / (L/g) | 0.00 |
| Objective Function | $2.16 \times 10^9$ |

Figure 3 presents two concentration profiles from the set of operating variables of $\xi_2$. Concentration profiles were plotted with TMB model optimal parameters and the screened Bi-Langmuir adsorption isotherm equation, where white squares and black dots are the measured experimental concentration of the more and less retained compound, respectively. The cloud in Figure 3 is the predicted values for enantiomers A|B. It is worth noting that, due to the randomness of the LHS, not all operating variables present a reasonable concentration profile as in Figure 3a. Some conditions present contaminations in the raffinate and extract streams, and, therefore, enantiomers A and B are not effectively separated, as shown in Figure 3b.

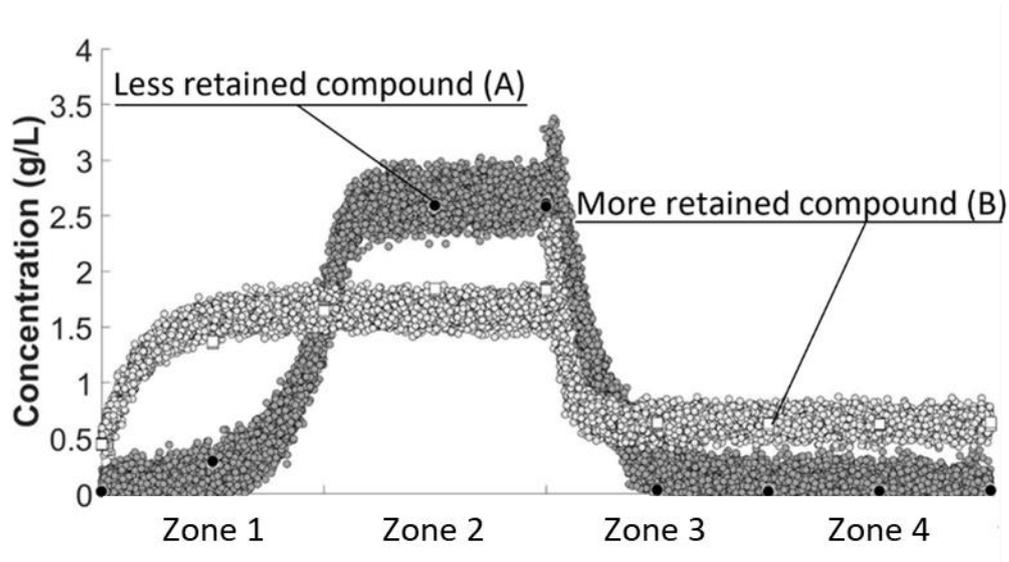

(a)

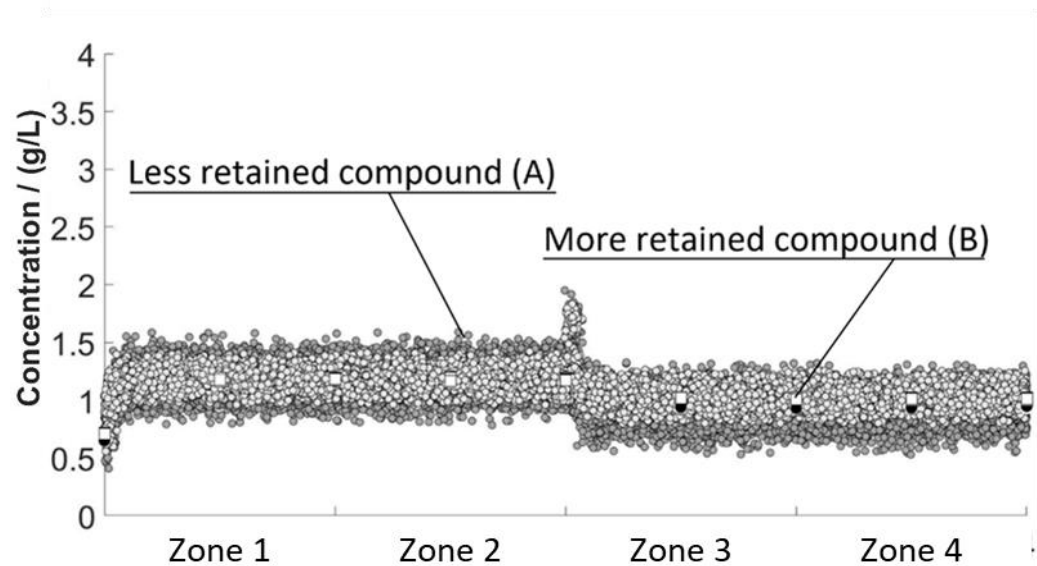

(b)

Figure 3: Two examples of operating conditions with experimental concentration points (white squares and black dots) and predicted concentration profiles (cloud) of TMB model with Linear adsorption isotherm equation whose parameters were estimated with 15-run series ($\xi_2$).

## 6.2. Analysis of adsorption isotherm equations

An analysis of the model prediction is carried out for $\xi_2$. Figure 4 shows the concentration profiles generated using test samples, i.e., two operating conditions taken from the literature,[7] shown in Table 7, considering the estimated parameters from Table 6. The predicted concentration profiles (cloud) of TMB model present good prediction capability.

Table 7: Operating conditions from literature.

|  | **Operating Condition 1** | **Operating Condition 2** |
|---|---|---|
| $c_{A|B}^F$ | $2.9\ g \cdot L^{-1}$ | $2.9\ g \cdot L^{-1}$ |
| $v_S^*$ | $12.16\ mL \cdot min^{-1}$ | $11.65\ mL \cdot min^{-1}$ |
| $v_{IV}^*$ | $27.27\ mL \cdot min^{-1}$ | $27.61\ mL \cdot min^{-1}$ |
| $v_E^*$ | $21.45\ mL \cdot min^{-1}$ | $21.45\ mL \cdot min^{-1}$ |
| $v_X^*$ | $17.98\ mL \cdot min^{-1}$ | $17.98\ mL \cdot min^{-1}$ |
| $v_F^*$ | $3.64\ mL \cdot min^{-1}$ | $3.64\ mL \cdot min^{-1}$ |

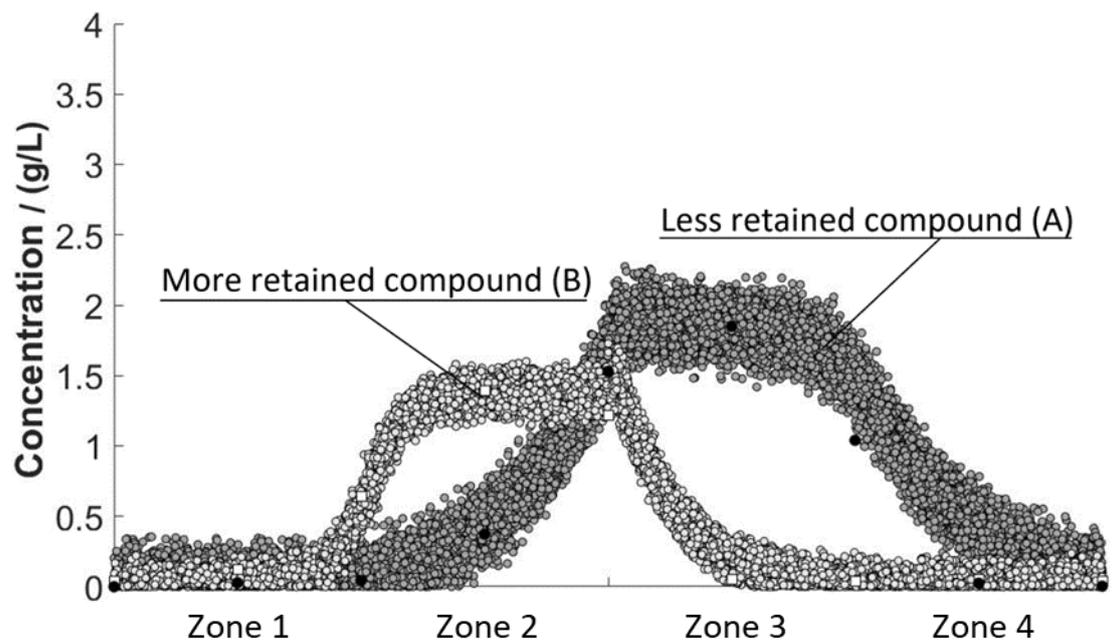

(a)

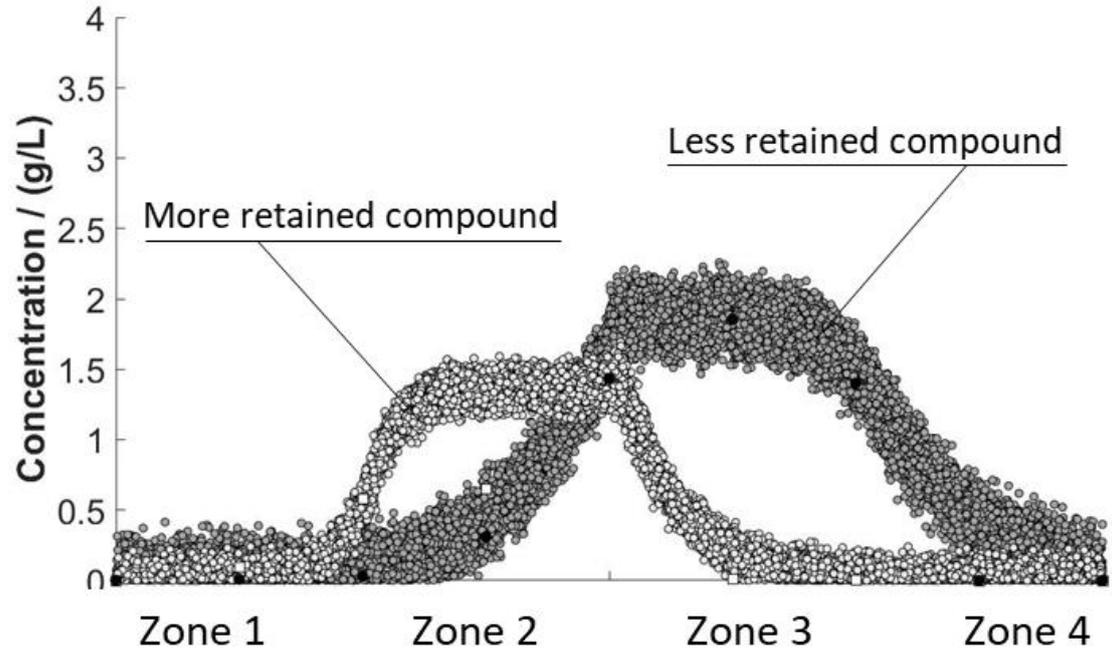

(b)

Figure 4: Experimental concentration points (squares and dots) and predicted concentration profiles (cloud) of TMB model with operating variables from the literature.

### 6.3. Uncertainty evaluation

The parameter confidence regions considering an $\alpha = 99.9\%$ are evaluated in Fisher-Snedecor ellipses in order to obtain a reliable model. A Bi-Langmuir adsorption isotherm equation for the experimental series $\xi_2$ suits better experimental and synthetic concentration points, and, therefore, its parameters ($Pe$, $\alpha$, $q_S^{(1)}$, $q_S^{(2)}$, $b_A^{(1)}$, $b_A^{(2)}$, $b_B^{(1)}$, and $b_B^{(2)}$) are chosen to be assessed in pairs in 2D and 3D plots of the Fisher-Snedecor, as shown in Figure 5. The objective function landscape for the PSO is represented in the 3D graphics. There is a convergence of the optimization and the topography in the surroundings towards the minimum point. The PSO particles hover over the landscape and concentrate around the optimal point.

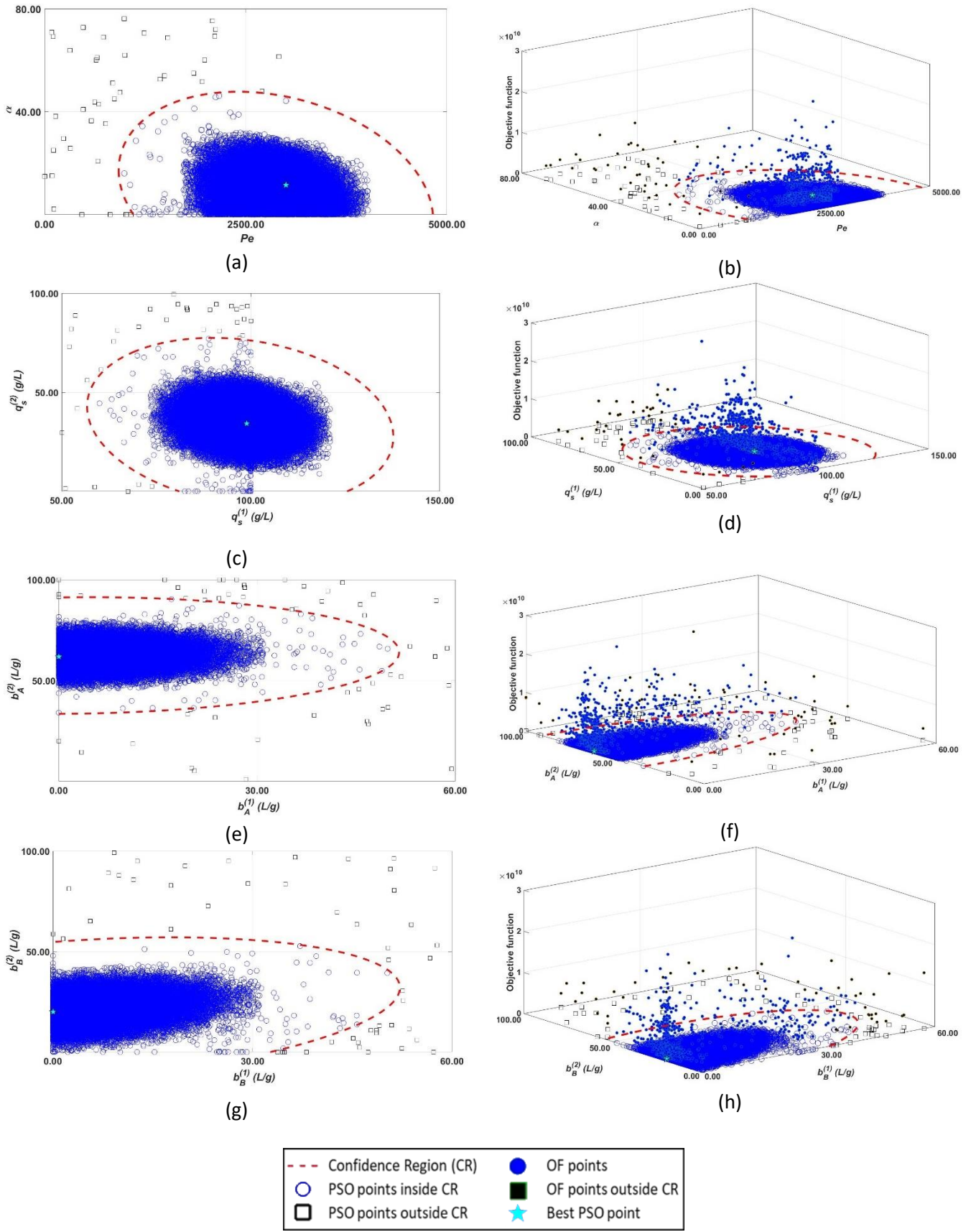

Figure 5: Fisher-Snedecor ellipses in two and three dimensions for $Pe$ and $\alpha$ (a) and (b); $q_S^{(1)}$ and $q_S^{(2)}$ (c) and (d); $b_A^{(1)}$ and $b_A^{(2)}$ (e) and (f) and $b_B^{(1)}$ and $b_B^{(2)}$ (g) and (h) parameters.

The correlation between parameters is somewhat significant. This can be observed by the angles between the ellipse and the axes of the confidence region between $Pe$ and $\alpha$; and $q_S^{(1)}$ and $q_S^{(2)}$ as shown in Figure 5. A steeper inclination is observed between $q_S^{(1)}$ and $q_S^{(2)}$ due to strong correlation between them.

## 7. Conclusion

In this work an extended global parameter estimation for the SMB phenomenological model is presented. It carries out a screening of adsorption isotherm in order to select the adsorption equation that better fits a system with little previous information. Design of experiments is carried out by LHS to generate well-dispersed spaced and diverse experimental operating variables for different series of experiments. This enables the determination of a minimum number of experiments necessary to obtain a representative model by parameter estimation.

Parameters are estimated with synthetic experimental data and predicted concentration values by comparison of a virtual plant and TMB model applying an SIL strategy. An analysis of the Bi-Langmuir isotherm equation with experimental data from literature shows a good prediction capability. The methodology shows that, in this case study, 15-run series present better results and the Bi-Langmuir isotherm equation fits betters in this scenario.

The results point to a total of 15 experiments as the optimal to obtain a precise model. It is a reasonable number of tests. However, the advantage of this approach is that the model is already identified within the system in contrast to the usual approach in SMB field. On the other hand, in the field of model identification, usually a model requires large dataset to be identified, which can be exhaustive. Hence, identifying the minimum number becomes an advantage.

From the results, a reasonable number of 15 experiments is optimal to obtain a precise model. The advantage of this approach is that the model is already identified within the system, in contrast to the usual approach in the SMB field. However, in the field of model identification, a large dataset is usually required to identify a model, which can be time-consuming. Therefore, identifying the minimum number is advantageous.

This work also demonstrates a correlation between parameters for the Bi-Langmuir isotherm equation in the 15-run series. However, this is already expected, due to the nature of the adsorption capacity of the parameters. Finally, this work contributes to SMB modeling literature, and it demonstrates that it is possible to evaluate different adsorption isotherm equations with minimal knowledge of the system.

## 8. Acknowledgments

This study was financed in part by the Coordenação de Aperfeiçoamento de Pessoal de Nível Superior - Brasil (CAPES) - Finance Code 001. This work was also financially supported by LA/P/0045/2020 (ALiCE), UIDB/50020/2020 and UIDP/50020/2020 (LSRE-LCM), funded by national funds through FCT/MCTES (PIDDAC). and FCT – Fundação para a Ciência e Tecnologia under CEEC Institutional program.

For Table of Contents Only

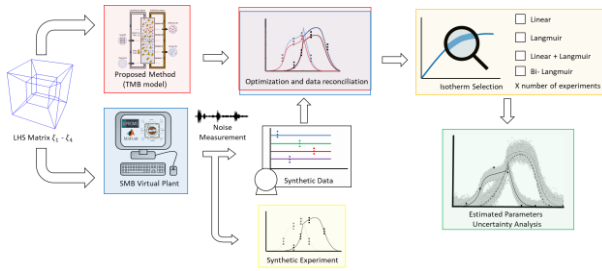